# Orientation Dependence of the Anomalous Hall Resistivity in Single Crystals of $Yb_{14}MnSb_{11}$

Brian C. Sales, Rongying Jin, and David Mandrus

Materials Science and Technology Division, Oak Ridge National Laboratory, Oak Ridge, Tennessee 37831

Abstract

The Hall resistivity, electrical resistivity and magnetization of single crystals of the tetragonal ferromagnet $Yb_{14}MnSb_{11}$ are reported as a function of the direction of the current, **I**, and magnetic field, **H** with respect to the principal crystallographic axes. With **I** along the unique *c* direction and **H** in the *a-b* plane, the anomalous Hall resistivity in the limit of zero applied field is negative for all temperatures $T < T_c = 53$ K. In this direction, the anomalous Hall effect behaves in a manner similar to that observed in other ferromagnets such as Fe, Co, $Mn_5Ge_3$, and $EuFe_4Sb_{12}$. However, with **I** in the *a-b* plane and **H** along the *c* direction, the anomalous Hall behavior is completely different. The anomalous Hall resistivity data are positive for all $T < T_c$ and a similar analysis of these data fails. In this direction, the anomalous response is not a simple linear function of the magnetization order parameter, and for a fixed temperature ($T < T_c$) does not depend on the magnitude of the magnetization perpendicular to the current in the *a-b* plane. That is, when the magnetization and applied field are rotated away from the *c* direction, the *anomalous Hall resistivity does not change*. In all other soft ferromagnets that we have examined (including La doped crystals of $Yb_{14}MnSb_{11}$, i.e. $Yb_{13.3}La_{0.7}MnSb_{11}$) rotation of the magnetization and magnetic field by an angle θ away from a direction perpendicular to **I** results in a decrease in both the anomalous and normal portions of the Hall resistivity that approximately scales as cos(θ). We suggest that the unique response exhibited by $Yb_{14}MnSb_{11}$ is a direct reflection of the delicate balance between carrier mediated ferromagnetism and Kondo screening.

75.30.Mb, 72.15.Qm, 75.30.-m

## Introduction

The compound $Yb_{14}MnSb_{11}$ is a magnetically soft, low carrier concentration ferromagnet with a Curie temperature, $T_c$, of 53 ±1 K. The compound was first synthesized by Chan, et al. [1], and the first single crystals were grown and characterized by Fisher et al. [2] X-ray absorption edge (XAS) and magnetic circular dichroism (XMCD) measurements[3] found no evidence of magnetism associated with Yb, and concluded that Yb has a nonmagnetic $Yb^{+2}$ configuration (filled 4f shell). This conclusion is supported by the observation of ferromagnetism at about 60 K in the isostructural compound $Ca_{14}MnSb_{11}$.[1] Low temperature magnetization measurements[2,4,] and XAS and XMCD data[3] suggest a $Mn^{+2}$ ($d^5$) configuration with the moment of one spin compensated by the anti-aligned spin of an Sb 5$p$ hole. This configuration is consistent with the observed saturation magnetization at 2 K of 4$\mu_B$ per Mn.[2,4] A $d^5$ + hole ($d^5 + h$) configuration is expected from electronic structure calculations on the related $Ca_{14}MnBi_{11}$ compounds[5] and a $d^5 + h$ configuration is also found[6] in the most heavily studied dilute magnetic semiconductor (DMS) GaAs:Mn. Good thermoelectric properties at elevated temperature have been reported recently for polycrystalline $Yb_{14}MnSb_{11}$ samples.[7]

Our original motivation for synthesizing large crystals of $Yb_{14}MnSb_{11}$ was to investigate the magnetism in a model DMS system. The ferromagnetic compound contains 3.8 at % Mn with nearly 1 carrier per Mn. Each Mn is at a well-defined crystallographic site in the structure with a minimum Mn-Mn separation of 1 nm. Unlike many of the DMS alloys investigated in the literature, clustering of magnetic ions does not occur in this compound. On the basis of optical and thermodynamic measurements, Burch et al. [8] were the first to suggest that the ferromagnetic ground state of $Yb_{14}MnSb_{11}$ is unusual. They proposed that $Yb_{14}MnSb_{11}$ is a rare example of an underscreened Kondo lattice, with a Kondo temperature $T_K \approx 300$ K. As the material is cooled from high temperatures, part of the entropy associated with each $Mn^{+2}$ *S*= 5/2 spin is removed via ferromagnetic ordering and part is transferred via hybridization with the itinerant Sb 5p states near the Fermi energy, resulting in a renormalization of the density of states and an increase in the carrier effective mass. Optical measurements[8] and heat capacity data down to 0.3 K[4], give values for the electronic specific coefficient, γ, of about 160 mJ/mole-$K^2$ and an effective mass, $m^* \approx 20\ m_e$. Seebeck and resistivity measurements under pressure and chemical doping studies are also consistent with an underscreened Kondo lattice ground state.[4, 9] This unusual ground state depends on a delicate balance between carrier mediated magnetic order and Kondo screening.[10, 11, 12, 13] Some unique aspects of the anomalous Hall data presented in this article may be related to this delicate balance.

Our initial Hall experiments on $Yb_{14}MnSb_{11}$ were aimed only at estimating the carrier concentration, not a study of the anomalous Hall effect. In the course of the Hall measurements, however, it became clear that there were some distinct advantages of studying the anomalous Hall effect in these types of low-carrier-concentration magnetically-soft ferromagnetic materials known as Zintl compounds.[14,15] The low carrier concentration ($\approx 10^{21}$ $cm^{-3}$) and small magnetic anisotropy make it easier to measure and separate the normal and anomalous Hall contributions on thinned single crystals with different crystallographic orientations. A preliminary report of the AHE in three such ferromagnetic compounds ($Yb_{14}MnSb_{11}$ [ *I* // *c*, *H*// *a*], $Eu_8Ga_{16}Ge_{30}$, and $EuFe_4Sb_{12}$) has been published.[16]

In the present article we focus on the orientation dependence of the AHE from single crystals of the tetragonal compound $Yb_{14}MnSb_{11}$. Because the AHE data from $Yb_{14}MnSb_{11}$ in one geometry ( *H* // *c*, *I*// *a*) is so unusual and unexpected, we also present AHE data from single crystals of two "normal" reference compounds $EuFe_4Sb_{12}$ ($T_c \approx 84$ K), and $Yb_{13.3}La_{0.7}MnSb_{11}$ ($T_c \approx 40$ K).

## Synthesis and Experimental Methods

Single crystals of $Yb_{14}MnSb_{11}$, $Yb_{13.3}La_{0.7}MnSb_{11}$, and $EuFe_4Sb_{12}$ are grown from molten metal fluxes.[17] The 14-1-11 crystals are grown from a Sn flux using a method similar to that reported by Fisher *et al.*[2] with initial molar compositions for Yb:Mn:Sb:Sn of 14:5:11:90, and Yb:La:Mn:Sb:Sn 12:2:5:11:90. If the Mn molar concentration is lowered to near 1, large crystals of another phase grew, namely $Yb_{11}Sb_{10}$:Mn or $Yb_{11}MnSb_9$. Single crystals of the filled skutterudite $EuFe_4Sb_{12}$ are grown from an Sb flux with a starting composition of Eu:Fe:Sb of 1:4:20 as described previously.[18] Energy dispersive X-ray analysis and X-ray structure refinements of the 14-1-11 crystals indicate that La only substituted for Yb, and that there was no evidence of antisite disorder; i.e., all of the Mn is confined to a unique crystallographic site and there is no mixing of Yb on Sb sites[4]. Structure refinement of the $EuFe_4Sb_{12}$ crystals indicates fewer than 5% of vacancies on the Eu site, in agreement with previous data.[18] The $EuFe_4Sb_{12}$ crystals grew as 3-5 mm size cubes, reflecting the underlying cubic crystal structure (*Im-3* $a$ = 0.917 nm, 34 atoms per conventional unit cell ). The 14-1-11 phase is tetragonal (*I41/acd* $a$ = 1.661 nm, $c$= 2.195 nm, 208 atoms per conventional unit cell), but the crystals have multiple faces and no simple growth habit (Fig 1). The La doped 14-1-11 crystals have a similar growth habit and lattice parameters ($a$ =1.661 nm, $c$ = 2.199 nm). Hall resistivity, electrical resistivity, and magnetoresistance measurements were made on thinned single crystal plates using the resistivity option, and horizontal rotator option for a physical property measurement system from Quantum Design. Six 0.025-mm diameter Pt wire leads were attached to each crystal using H20E silver epoxy from EpoTek. The 14-1-11 crystals were lightly sanded before attaching the leads. The epoxy was cured at 373 K in a nitrogen atmosphere to avoid oxidation of the sample surface. Typical contact resistance was less than 5 ohms. For the Hall resistivity measurements, any voltages due to a mis-position of the Hall leads were corrected by either reversing the direction of the magnetic field or by rotating the crystal by 180 ° in a fixed magnetic field. As in any Hall measurement, only voltages that are an odd function of the magnetic field were kept. Magnetization data were collected using a commercial superconducting quantum interference device (SQUID) magnetometer from Quantum Design.

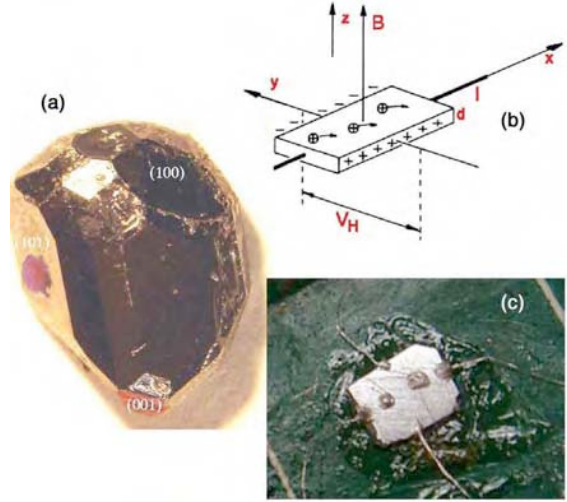

*FIG. 1. (Color Online) a) Crystal of $Yb_{14}MnSb_{11}$ illustrating the unusual growth habit. Several of the facets are identified using x-ray diffraction. As-grown the crystal weighed 0.67 g. (b) A schematic of the geometry used for the Hall measurements. In the rotation experiments the sample is tilted about the y axis. (c) Hall and resistivity leads attached to a crystal that is thinned and ground into a rectangular plate measuring about 3 mm x 4mm x 0.7 mm and weighing 80 mg.*

## Results

Typical Hall resistivity data from single crystals of $Yb_{14}MnSb_{11}$ in two orientations are shown in Fig. 2. The data are from two different crystals that were oriented and polished into a thin plate. Three crystals in each orientation were examined to ensure that the results were due to the orientation of the crystal and not to the slight variations in properties among crystals from different growth batches. For a magnetic material the Hall resistivity is normally described by $\rho_{xy} = R_0 B + \rho_{xy}'$ where $R_0 B$ is the ordinary contribution and $\rho_{xy}'$ describes the anomalous contribution to the Hall effect (AHE). The Hall coefficient, $R_0$, is inversely proportional to the carrier concentration in simple doped semiconductors. In many materials $\rho_{xy}'$ is proportional to the magnetization, M, and is often written as $\rho_{xy}' \approx R_s 4\pi M$.[19,20,21] In general, however, $\rho_{xy}'$ is a more complicated function of M that can in principle be determined from the electronic structure.[22,23,24,25,26,27,28,29] In the limit of *zero applied magnetic field*, $\rho_{xy}'$ can be parameterized

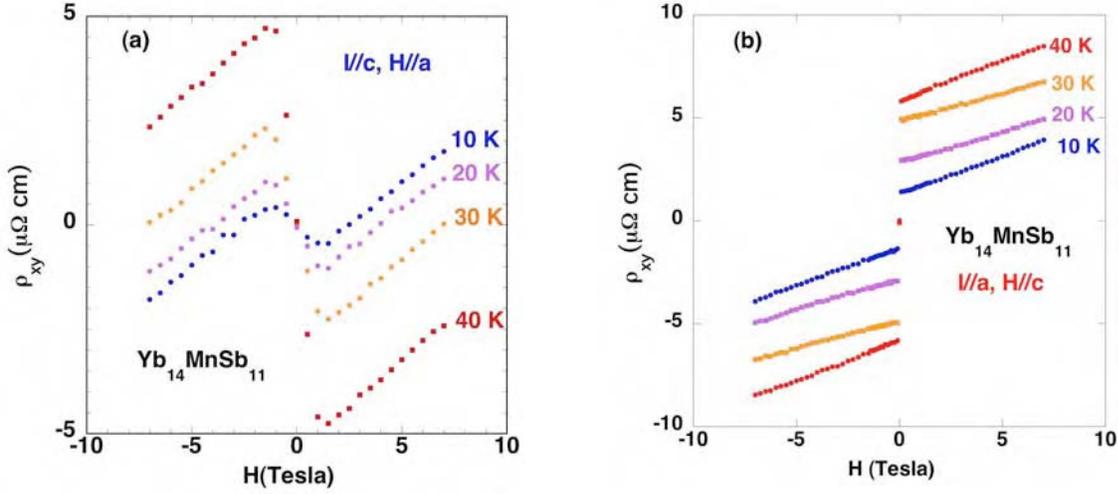

*FIG. 2. (Color online) (a) Hall data from a $Yb_{14}MnSb_{11}$ crystal with H//a and I//c. For clarity only a portion of the Hall data are shown. For positive values of magnetic field (H > 2 Tesla) the extrapolated intercepts at H = 0 are negative. The extrapolated intercept at H=0 is defined as $\rho'_{xy}$, the anomalous Hall resistivity in the limit of zero applied magnetic field. These data are similar to those reported previously by us on another $Yb_{14}MnSb_{11}$ crystal with I//c and H// (110) direction (see Fig 8, ref 16). (b) Hall data from a $Yb_{14}MnSb_{11}$ crystal with H//c and I//a. For clarity only a portion of the Hall data are shown. For positive values of magnetic field (H > 2 Tesla) the extrapolated intercepts at H = 0 are positive, opposite to that found in Fig. 2a.*

[30,16,21] by $\rho'_{xy} = [\sigma'_{xy} \rho^2 + a \rho] f[M(T)/M(0)]$, where $\rho$ is the zero field resistivity $\rho_{xx}(H \approx 0)$, $\sigma'_{xy}$ is the intrinsic anomalous Hall conductivity, $a$ describes the extrinsic contribution from skew scattering, and $f$ is a general function of the spontaneous magnetization M(T). At low temperatures (T<< $T_c$), the spontaneous magnetization saturates and $f[M(T)/M(0)] \approx 1$ while for T>$T_c$, $f[M(T)/M(0)] = 0$. To be consistent with the notation used in most theoretical calculations we set $\sigma'_{xy} = \sigma_{xy} \approx \sigma^0_{xy}$ where $\sigma^0_{xy}$ is the anomalous Hall conductivity in the limit T << $T_c$. The limited theoretical calculations[21,26] to date on a variety of different materials indicate $\sigma_{xy}$ is at most a very weak function of temperature and hence we approximate it by its low temperature value $\sigma^0_{xy}$. The skew scattering coefficient, $a$, is also assumed to be independent of temperature. With these approximations the anomalous Hall resistivity in the *limit of zero applied magnetic field* is described by $\rho'_{xy} = [\sigma^0_{xy} \rho^2 + a^0 \rho] [M(T)/M(0)]$, where we have also assumed the simplest approximation for $f$, i.e. $f[M(T)/M(0)] = M(T)/M(0)$. The experimental value of $\rho'_{xy}$ is determined by extrapolating the Hall resistivity data,

such as shown in Fig. 2, from higher magnetic fields back to H= 0. Since only small magnetic fields are required to align the magnetic domains, and since the normal portion of the Hall resistivity is rather large, this extrapolation is straightforward. However, we note that in some materials, such as MnSi, this procedure is complicated by a large magnetoresistance and a high carrier concentration.[31]

The most obvious difference in the data shown in Figs 2a and 2b is a change in the sign of $\rho_{xy}'$. In Fig 2a (**H**//*a*, **I**//*c*) $\rho_{xy}'$ is negative for all temperatures while for **H**//*c*, **I**//*a*, $\rho_{xy}'$ is positive for all temperatures (Fig 2b). To our knowledge, this is the first example in which the sign of the AHE depends on the crystallographic orientation. The sign change persists in finite magnetic fields to temperatures well above $T_c$, as illustrated in Fig 3. The carrier concentration for each crystal is estimated from Hall resistivity data at 5 K to be $1.6 \pm 0.1 \times 10^{21}$ holes/cm$^3$, which corresponds to about 1 hole per Mn. These values are similar to those reported by us previously[16], and to the high temperature values reported by Brown *et al.*[7] for a polycrystalline sample.

The anomalous Hall resistivity in the *limit of zero applied magnetic field* is analyzed using $\rho_{xy}' = [\sigma^0_{xy} \rho^2 + a^0 \rho] [M(T)/M(0)]$. The experimental quantities $\rho_{xy}'$, $\rho$, and $M(T)/M(0)$ were measured at each temperature $T < T_c$. The intrinsic Hall conductivity, $\sigma^0_{xy}$ and the skew scattering coefficient $a^0$ are determined by fitting a line to plots of $[M(0)/M(T)] \rho_{xy}'/\rho$ versus $\rho$. The slope of the line yields $\sigma^0_{xy}$ and the intercept $a^0$. Using the data shown in Fig 2a, along with the measured resistivity and spontaneous magnetization data (see Reference 16 for more details), results in values of $\sigma^0_{xy} = -32 \pm 1$ $\Omega^{-1}$ cm$^{-1}$ and $a^0 = -0.0037$ (Fig. 4a). These values have the same sign and magnitude found before with **H**// (110), **I**//*c* ($\sigma^0_{xy} = -14 \pm 1$ $\Omega^{-1}$ cm$^{-1}$, -0.0033). In the present case the value of $\sigma^0_{xy}$ is approximately twice as large as found previously[16], probably reflecting both the variation of $\sigma^0_{xy}$ with the direction of **H** in the basal plane and the inevitable slight variation in properties (residual resistivity for example) among crystals from different growth batches. Analysis of the Hall data with **H**//*c* and **I**//a, however, results in qualitatively different behavior unlike any we have observed in our investigations of a variety of ferromagnets. Using data, part of which is displayed in Fig. 2b, and assuming, as before, that $\rho_{xy}' = [\sigma^0_{xy} \rho^2 + a^0 \rho][M(T)/M(0)]$, results in the analysis displayed in Fig. 4b. This analysis of the AHE clearly fails for this crystal orientation. Similar behavior was found on two other $Yb_{14}MnSb_{11}$ crystals with the same Hall geometry. One or more of the assumptions used to parameterize $\rho_{xy}'$ is clearly not valid for this orientation of the crystal.

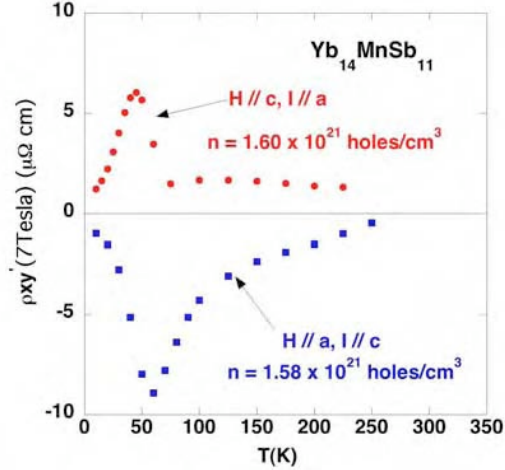

*FIG. 3. (Color Online) Anomalous Hall resistivity at 7 Tesla vs temperature. The carrier concentration at 5 K was estimated for each crystal and assumed to be approximately constant at higher temperatures.*

To help understand the unusual behavior shown in Fig. 4b, we examined the AHE from chemically doped crystals of $Yb_{14}MnSb_{11}$, namely $Yb_{13.3}La_{0.7}MnSb_{11}$. The replacement of $Yb^{+2}$ by $La^{+3}$ ions tends to reduce the hole concentration, introduce some chemical disorder and reduce the screening of the $Mn^{+2}$ moments.[4] These crystals are ferromagnetic with $T_c \approx 42 \pm 2$ K, depending on the exact growth conditions, have a lower carrier concentration, a higher saturation magnetization, and a much higher residual resistivity for $T << T_c$ (Fig 5). Because of a smaller change in $\rho$ below $T_c$ (Fig 5), the Hall resistivity does not vary much with temperature (Fig 6a). Analysis of these data (Fig 6b) from the La doped crystals with **H**//*c*, and **I**//*a*, however, reveals a conventional

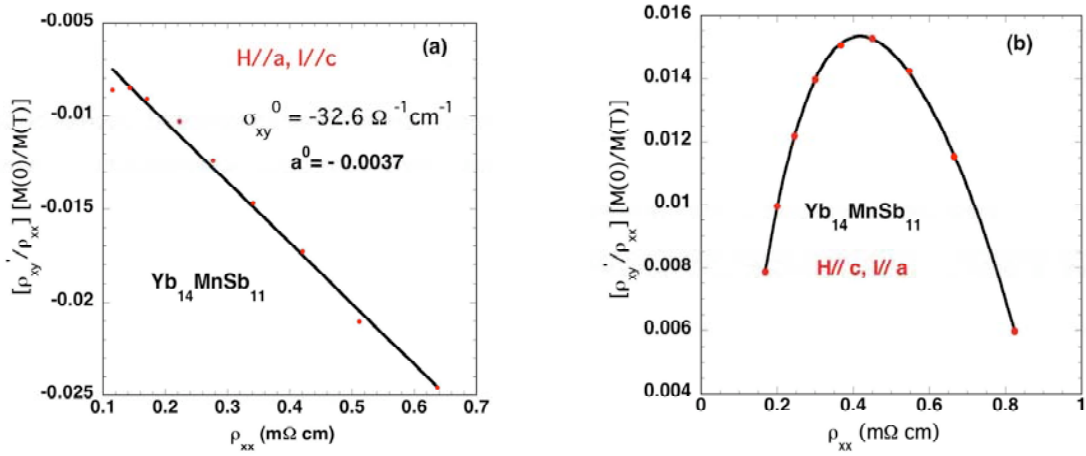

*FIG. 4. (a) Analysis of the intrinsic and extrinsic contributions to the AHE of an $Yb_{14}MnSb_{11}$ crystal with $H//a$ and $I//c$. The approximation of $\rho_{xy}' = [\sigma^0_{xy} \rho^2 + a^0 \rho] [M(T)/M(0)]$ is used to analyze the data. (b) Attempted analysis of the intrinsic and extrinsic contributions to the AHE of an $Yb_{14}MnSb_{11}$ crystal with $H//c$ and $I//a$.*

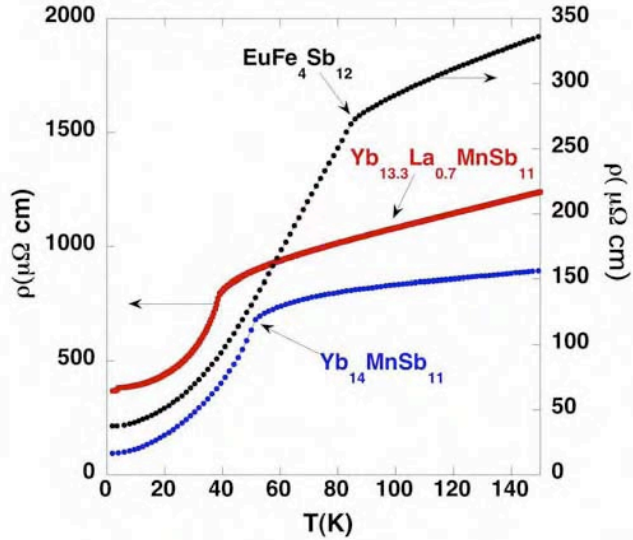

*FIG. 5. (Color Online) Resistivity versus temperature for three ferromagnets. The left resistivity scale is for the La-doped and undoped 14-1-11 crystals. Note the relatively small decrease in the resistivity of the La-doped crystal below $T_c \approx 40$ K. The resistivity data for each crystal is recorded during the Hall measurements using two additional leads near the center of the rectangular plate (see Fig 1c). Although this is not the best geometry for determining the absolute values of the resistivity, this method eliminates variations in resistivity among different crystals. The absolute values of the resistivity are only accurate to within about 20%,*

*mainly due to the uncertainty in the geometry of the leads. The resistivity data in the figure is associated with the Hall data displayed in Figs. 4a, 6 and 11.*

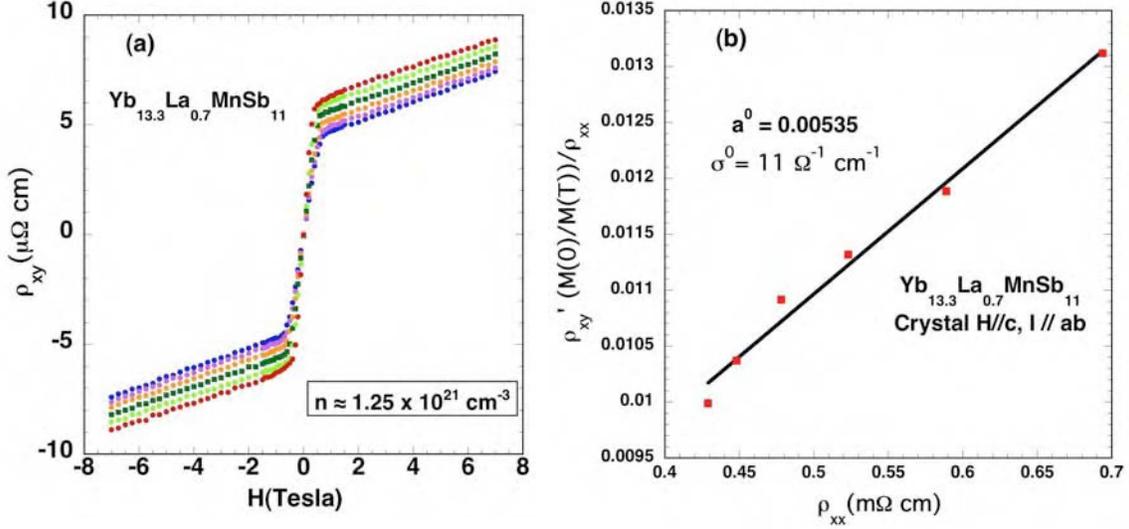

*FIG. 6. (Color Online) (a) Hall resistivity vs field with H//c, and I//a for a La doped crystal for temperatures between 35 K and 10 K. (b) Analysis of the intrinsic and extrinsic contributions to the AHE using the approximation of $\rho_{xy}' = [\sigma^0_{xy} \rho^2 + a^0 \rho] [M(T)/M(0)]$ to analyze these data.*

dependence of the AHE on temperature and resistivity, in sharp contrast to the behavior exhibited by the undoped crystals (Fig 4b). Hall data from the La doped crystals with **H**//**a** and **I**//**c** (not shown) exhibit negative values of $\rho_{xy}'$ indicating a change in sign of the AHE with crystal orientation for both the doped and undoped crystals.

Single crystals of the tetragonal ferromagnets $Yb_{14}MnSb_{11}$ and $Yb_{13.3}La_{0.7}MnSb_{11}$ both exhibit a change in sign of the anomalous Hall resistivity, $\rho_{xy}'$, depending on the orientation of the crystal. With **H**//*c* and **I**//*a*, $\rho_{xy}'$ is positive, while when **H**//*a* and **I**//*c*, $\rho_{xy}'$ is negative. The temperature dependence of the anomalous Hall resistivity behaves normally except for the undoped crystal with **H**//*c* and **I**//*a* (see Fig. 4b). To explore this unusual behavior, we examined the effects of tilting the sample with respect to the magnetic field on the measured Hall resistivity. Referring to Fig 1b, the sample was rotated around the **y** axis by an angle Θ. The Hall voltage is corrected for any offset of the Hall leads by reversing the applied magnetic field and only keeping voltages that are odd in the magnetic field. These crystals are soft ferromagnets which means that for fields larger than about 2 Tesla, the applied magnetic field **H** and the magnetization **M** of the sample point in the same direction. This is illustrated in Fig. 7 where the magnetization of one of the undoped $Yb_{14}MnSb_{11}$ crystals is shown as a function of the angle between the easy **c** axis and the applied magnetic field. Similar magnetization results are found for the La doped crystals. As the crystal is rotated around the **y** axis by an angle $\Theta$, the component of **H** perpendicular to the current *I* is H cos $\Theta$, and if **H** > 2 Tesla, the component of **M** perpendicular to *I* is M cos $\Theta$.

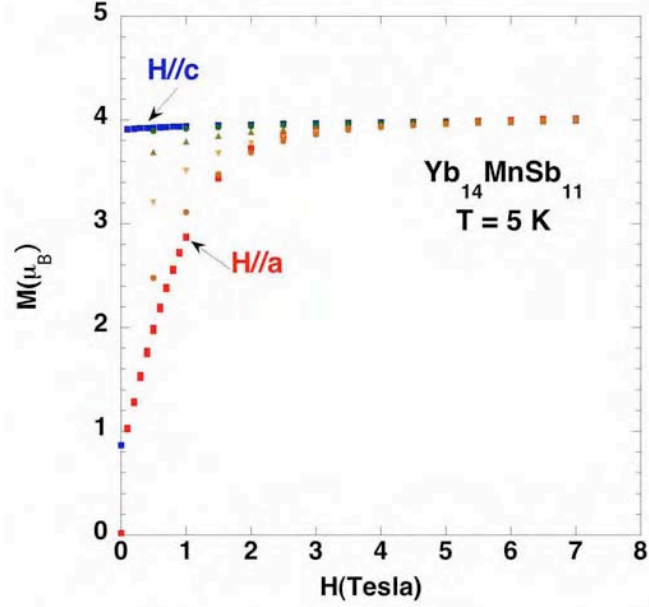

*FIG. 7. (Color Online) Magnetization versus applied magnetic field. Data are shown with H//c, and with the sample rotated toward the a axis by 20°, 40°, 60°, 80° and 90° (H//a). For fields larger than about 2 Tesla, the sample magnetization M is aligned along the direction of the applied magnetic field H.*

Typical Hall resistivity data below $T_c$ for a $Yb_{14}MnSb_{11}$ crystal that is rotated with respect to the applied magnetic field is shown in Fig. 8a. With 0 degree rotation **H**//**c** and *I*//*a*. As the sample is rotated about the **y** axis by $\Theta$, the slope of the Hall resistivity data in finite magnetic fields decreases approximately as cos $\Theta$ (see Fig 8b). This is expected since the slope is primarily due to the normal Hall effect and the Lorentz force depends on the component of **B** perpendicular to *I*, i.e. B cos $\Theta \approx$ H cos $\Theta$ (For the 14-1-11 crystals $4\pi M_{sat}$ is only 600 gauss) . The value of the anomalous Hall resistivity, $\rho_{xy}'$, however, does not depend on the angle of rotation for rotation angles up to at least 80° !

This result is so unexpected, that we repeated the measurements on 2 other $Yb_{14}MnSb_{11}$ crystals with the same orientation (**H**//*c* and **I**//*a*) and for various temperatures well below $T_c$. The conclusion is the same, in this orientation $\rho_{xy}'$ does not depend on the angle of rotation. Since **M** essentially follows the direction of the applied field (see Fig. 7), this means that within our experimental resolution, $\rho_{xy}'$ does not depend on the direction of **M**. To illustrate how unusual and unexpected this result is, we present several examples of "normal" behavior.

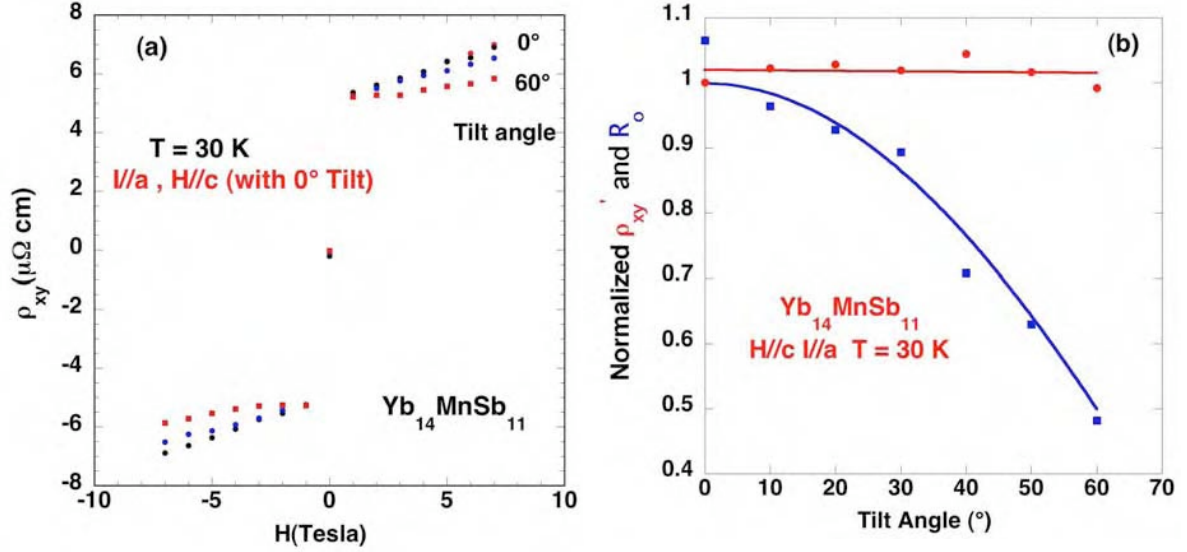

*FIG. 8. (Color Online) (a) Hall resistivity data at 30 K from a $Yb_{14}MnSb_{11}$ for various rotation angles about the y axis (see Fig. 1b). The crystal is prepared with I//a and H//c for 0 tilt angle. Note that $\rho_{xy}'$ is essentially independent of tilt angle. (b) Analysis of Hall resistivity data, a portion of which is displayed in Fig. 8a. The slope of the Hall resistivity data for H > 0.5 Tesla is defined as $R_0$. The variation $R_0$ with tilt angle is fit to $R_N \cos \Theta$. Squares are $R_0/R_N$, and blue line is $\cos \Theta$. The normalized anomalous Hall resistivity is defined as $\rho_{xy}'(\Theta)/\rho_{xy}'(0)$. The red line is a least squares fit to the AHE data.*

The same rotation experiments were performed on an undoped $Yb_{14}MnSb_{11}$ crystal but with **H**//*a* and **I**//*c* (crystal used for Figs. 2a and 4a), a La doped crystal with **H**//*c* and **I**//*a* (crystal used for Fig. 6), and an $EuFe_4Sb_{12}$ crystal (resistivity data shown in Fig. 5). Some results from these experiments are displayed in Figs. 9, 10 and 11. For all three crystals $\rho_{xy}'$ decreases approximately as $\cos(\Theta)$ when the applied magnetic field and magnetization are rotated by an angle $\Theta$ with respect to the direction of the current through the crystal.

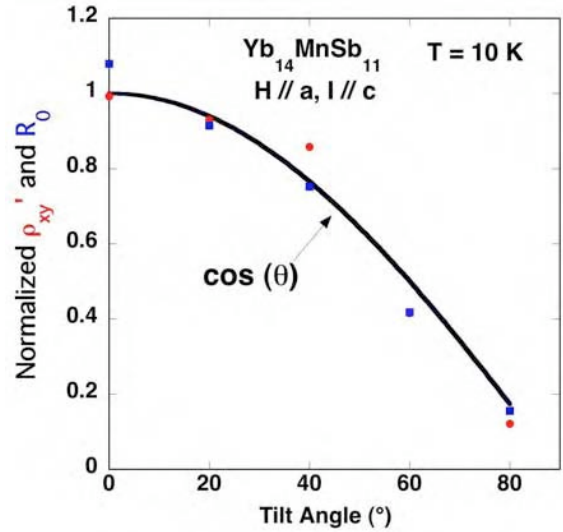

*FIG. 9. (Color Online) Effect of the rotation of a $Yb_{14}MnSb_{11}$ crystal (I//c, and H//a for 0 tilt) about the y axis on the normal and anomalous contributions to the Hall resistivity. Both contributions approximately decrease as $\cos(\Theta)$.*

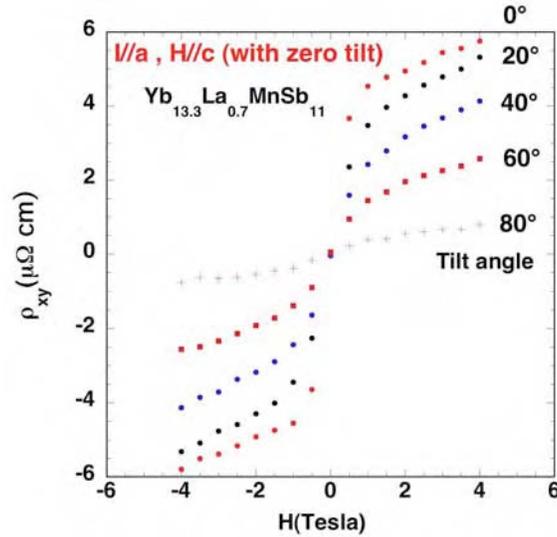

*FIG. 10 . Hall resistivity data from a La doped $Yb_{14}MnSb_{11}$ crystal at 5 K with I//a and H//c for zero tilt angle. Note the both the intercept and the slope of the data for H > 2 Tesla decrease as the sample is rotated about the y axis (see Fig 1b) in contrast to the response of the pure $Yb_{14}MnSb_{11}$ crystal with the same orientation (see Fig. 8a).*

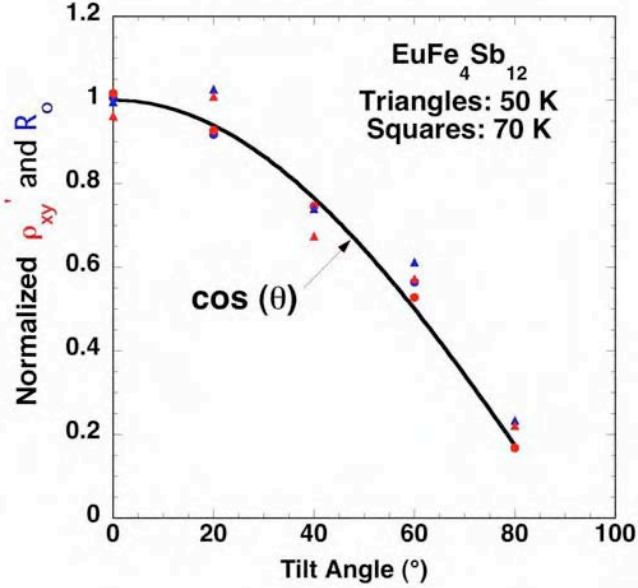

*FIG. 11. Effect of the rotation of a cubic EuFe$_4$Sb$_{12}$ crystal (I//(100), and H//(001) for 0 tilt) about the (010) axis on the normal and anomalous contributions to the Hall resistivity. Both contributions approximately decrease as cos(Θ). Hall resistivity data from two temperatures are shown.*

## Discussion

The anomalous Hall resistivity of ferromagnets, $\rho_{xy}'$ is usually described as originating from two sources: an extrinsic contribution due to skew scattering[32,33] that is proportional $\rho$ and an intrinsic contribution[19,22] that is proportional to $\rho^2$. A contribution to $\rho_{xy}'$ proportional to $\rho^2$ is intrinsic in the sense that the more fundamental Hall conductivity, $\sigma_{xy} \approx \rho_{xy}/\rho^2$ (just from inverting the resistivity tensor) becomes independent of scattering. The seminal work of Karplus and Luttinger[22] showed that the intrinsic contribution originated from the spin-orbit coupling of Bloch bands, which in principle could be determined from careful calculations of the electronic structure. Berry[34] showed that this type of intrinsic contribution is a very general consequence of the application of quantum mechanics to condensed matter systems that do not have either time reversal invariance (a ferromagnet for example) or inversion symmetry (many crystal structures lack inversion symmetry). The insight provided by Berry on the consequence of Berry phase effects on Bloch electrons in condensed matter systems has stimulated several groups to calculate the intrinsic anomalous Hall conductivity both at T=0 and as a function of temperature. Calculations of the intrinsic Hall conductivity from simple ferromagnets[26], such as Fe, have shown that the sign and the magnitude of $\sigma_{xy}$ are dominated by "hot spots" in the electronic structure and that there is no simple relationship between $\sigma_{xy}$ and the sign of the normal Hall coefficient. It is then perhaps not too surprising that for a complex tetragonal ferromagnet such as $Yb_{14}MnSb_{11}$ with 208 atoms in the conventional unit cell, that the sign of $\sigma_{xy}$ is different depending on whether the current is along *c* or *a*. Although detailed calculations of the intrinsic Hall conductivity are probably not feasible at the present time for $Yb_{14}MnSb_{11}$, the approximate magnitude[35] of $\sigma_{xy}$ should be given by $|\sigma_{xy}| \approx 0.1\ e^2\ k_F/h \approx 100\ \Omega^{-1}\ cm^{-1}$, as compared with the measured value of $|\sigma_{xy}| = 33\ \Omega^{-1}\ cm^{-1}$ (see Fig 4a). The data in Fig 4b are more difficult to understand. While a sign change with crystal direction can be rationalized within the Berry scenario of the AHE, the completely different functional dependence of $\rho_{xy}'$ on $\rho$ and M when **I//a** and **H//c** is puzzling since the resistivity is not very anisotropic (i.e. $\rho_c/\rho_a \approx 1.4$), the magnetic susceptibility well above $T_c$ is isotropic, and the ferromagnetism is fairly soft. In this same geometry (***I//a***), if the direction of the **M** is rotated by an angle $\Theta$ such that the component of **M** perpendicular to ***I*** is **M** $cos(\Theta)$, $\rho_{xy}'$ *remains unchanged* (see Fig. 8). This means that in this geometry $\rho_{xy}'$ *does not depend on the direction of* **M**. The expected behavior for $\rho_{xy}'$ is recovered, however. if a small amount of the non-magnetic $Yb^{+2}$ is replaced by non-magnetic $La^{+3}$ (i.e. $Yb_{13.3}La_{0.7}MnSb_{11}$). A small amount of chemical alloying is apparently enough to destroy the unique features of the ground state observed in the undoped $Yb_{14}MnSb_{11}$ compound. As discussed in the introduction, there is strong evidence that $Yb_{14}MnSb_{11}$ is a rare example of an underscreened Kondo lattice with a Kondo temperature $T_K \approx 300$ K.[4,8] This unusual ground state depends on a delicate balance between carrier mediated magnetic order and Kondo screening. It seems likely that some of the unique aspects of the anomalous Hall data presented in Figs. 4b and 8 are related to the interplay between Kondo physics and ferromagnetism. Since the data shown in Fig 4b and 8 indicate a different dependence of $\rho_{xy}'$ on $\rho$ and M, the data in Fig. 4b are replotted in Fig 12a and fit to a polynomial of the form $\rho_{xy}' = a + b\ \rho + c\ \rho^2$. A good fit to the data is obtained with a = -4.33 $\mu\Omega$ cm, b = 0.0389 and c = -37.5 $\Omega^{-1}$ cm$^{-1}$. The coefficient of the $\rho^2$ term is quite close to the value obtained for $\sigma_{xy}$ from the "normal" crystal orientation (Fig 4a). This suggests that when *I*//*a* and **H**//*c* there is an additional large and positive contribution to $\rho_{xy}'$ that is not present when *I*//*c* and **H**//*a*. The additional contribution is linear in $\rho$ with an offset of $-4.33$ $\mu\Omega$ cm as shown in Fig 12b. If a residual resistivity, $\rho_o$, of 110 $\mu\Omega$ cm is subtracted from the resistivity, the additional contribution is linear in ($\rho - \rho_o$). In Heavy Fermion or Kondo-lattice compounds[36,37], both theory and experiment show a large

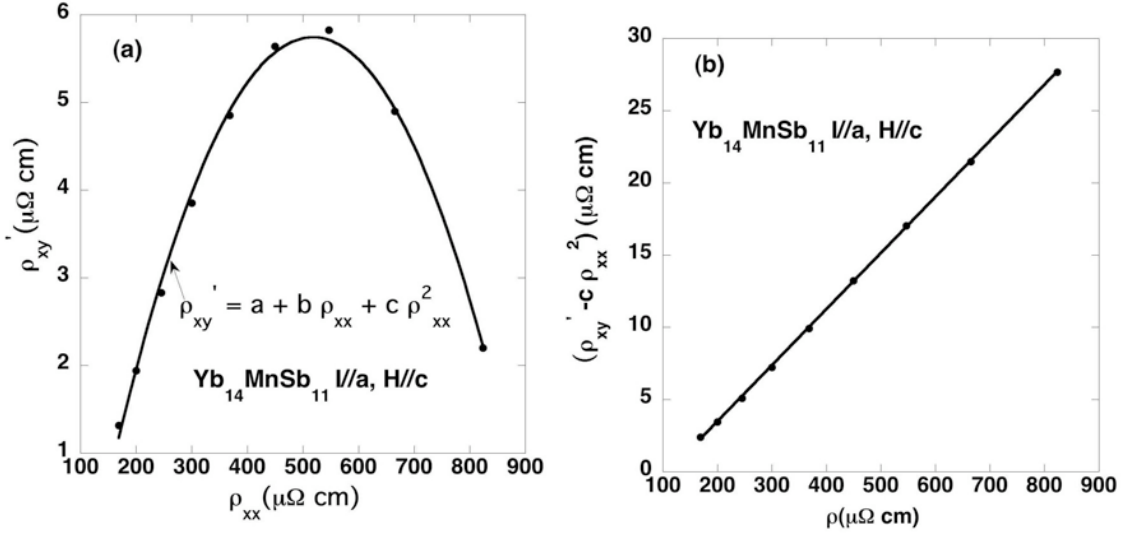

*Fig. 12 (a) AHE effect data shown in Fig 4b replotted as $\rho_{xy}'$ vs $\rho_{xx}$. A simple polynomial in $\rho_{xx}$ accurately describes the data with coefficients a = -4.334 $\mu\Omega$ cm, b = 0.0389 and c = -37.5 $\Omega^{-1}$ cm$^{-1}$. Note that the coefficient of $\rho_{xx}^2$ is surprisingly close to the value of – 32.6 $\Omega^{-1}$ cm$^{-1}$ obtained from the $Yb_{14}MnSb_{11}$ crystal in Fig4a. (b) Data in 12a with -37.5 $\rho_{xx}^2$ subtracted from measured $\rho_{xy}'$ values versus $\rho_{xx}$. The linear term approaches zero for $\rho_{xx} \approx 110$ $\mu\Omega$ cm, close to the residual resistivity value for this crystal. The large positive contribution to $\rho_{xy}'$ proportional to ($\rho_{xx}-\rho_0$) may be due to enhanced skew scattering associated with the Kondo effect.*

contribution to the Hall resistivity that is proportional to ($\rho-\rho_o$) times the magnetic susceptibility $\chi$ for temperatures well below $T_k$ ($T_k \approx 300$ K for $Yb_{14}MnSb_{11}$)[4,6]. For a Kondo lattice system well below $T_k$ the susceptibility is essentially independent of temperature, which would make the portion of the Hall resistivity due to the Kondo effect proportional to ($\rho-\rho_o$). In normal Kondo-lattice compounds, (such as $UPt_3$)[36] of course $\rho_{xy}$ and $\rho_{xy}'$ are zero in the limit H = 0, but for $Yb_{14}MnSb_{11}$, which is also ferromagnetic, the internal magnetic field that develops below $T_c \approx$ 53 K may result in an observable Kondo-like contribution to $\rho_{xy}'$ that is linear in $(\rho-\rho_o)$. This is a plausible explanation for the large additional contribution linear in $(\rho-\rho_o)$ that is observed in Fig 12. The strong skew scattering due to the Kondo effect appears to be much stronger when *I* is in the a-b plane. Strong anisotropic skew scattering is consistent with the theoretical calculations of Sanchez-Portal *et al.*[5] who find highly directional magnetic coupling between Mn ions in a related ferromagnetic compound $Ca_{14}MnSb_{11}$. Skew scattering dominates $\rho_{xy}'$ of $Yb_{14}MnSb_{11}$ when *I//a* and *H*//c. Since in "normal" Kondo lattice compounds the skew scattering contribution is proportional to the susceptibility (rather than the magnetization), it may be that the effective internal magnetic field only has to be larger than a critical value for Kondo skew scattering to be effective. This might provide a qualitative explanation of the results from the tilting experiments shown in Fig 8. Clearly, however, further theoretical insight is needed before one can claim that the Hall data from $Yb_{14}MnSb_{11}$ is understood.

## Conclusions

The tetragonal compound $Yb_{14}MnSb_{11}$ is a rare example of an underscreened Kondo ferromagnet[4,8,9] with $T_k \approx$300 K and $T_c \approx$53 K. Each $Mn^{+2}$ is at a well defined crystallographic site with a $d^5$ + *hole* electronic configuration[5] similar to that found in GaAs:Mn[6]. For $T<T_c$ the anomalous Hall resistivity, $\rho_{xy}'$, in the limit of zero applied magnetic field is negative when **H**//*a* and *I*//*c*, and positive when **H**//*c* and *I*//*a*. In both orientations analysis of $\rho_{xy}'$ yields an intrinsic Hall conductivity of $\sigma_{xy}^0 \approx -35$ $\Omega^{-1}$ $cm^{-1}$ but the skew scattering contributions differ by an order of magnitude : $a^0$ = -0.0037 when **H**//*a*, *I*//*c*, and $a^0$ = 0.0389 when **H**//*c*, *I*//*a*. The gigantic skew scattering observed with *H*//c is likely due to resonant Kondo scattering, although why it is so large only in one orientation is not completely understood. When skew scattering dominates (*I//a* ), rotation of **M** from **M**//*c* to **M** ≈// **I**//**a** does not change $\rho_{xy}'$ implying that $\rho_{xy}'$ does not depend on the direction of **M**.

## Acknowledgements

It is a pleasure to acknowledge stimulating and illuminating discussions with Allan MacDonald, Qian Niu and Peter Khalifah. Research sponsored by the Division of Materials Sciences and Engineering,Office of Basic Energy Sciences, U.S. Department of Energy, under contract DE-AC05-00OR22725 with Oak Ridge NationalLaboratory, managed and operated by UT-Battelle, LLC.